\documentclass[prodmode,acmtomccap]{acmlarge}

\usepackage{epstopdf}
\usepackage[bookmarksnumbered, colorlinks]{hyperref}

\usepackage{amsmath} 

\usepackage[english]{babel}
\usepackage[autostyle]{csquotes}

\usepackage{subfig} 
\acmVolume{2}
\acmNumber{3}
\acmArticle{1}
\articleSeq{1}
\acmYear{2013}
\acmMonth{8}

\usepackage[ruled, vlined]{algorithm2e}
\SetAlFnt{\algofont}
\SetAlCapFnt{\algofont}
\SetAlCapNameFnt{\algofont}
\SetAlCapHSkip{0pt}
\IncMargin{-\parindent}

\title{Speaker-following Video Subtitles}

\author{YONGTAO HU \affil{The University of Hong Kong}
JAN KAUTZ \affil{University College London}
YIZHOU YU \affil{The University of Hong Kong}
WENPING WANG \affil{The University of Hong Kong}
}

\begin{abstract}
We propose a new method for improving the presentation of subtitles in video (e.g. TV and movies). With conventional subtitles, the viewer has to constantly look away from the main viewing area to read the subtitles at the bottom of the screen, which disrupts the viewing experience and causes unnecessary eyestrain. Our method places on-screen subtitles next to the respective speakers to allow the viewer to follow the visual content while simultaneously reading the subtitles. We use novel identification algorithms to detect the speakers based on audio and visual information. Then the placement of the subtitles is determined using global optimization. A comprehensive usability study indicated that our subtitle placement method outperformed both conventional fixed-position subtitling and another previous dynamic subtitling method in terms of enhancing the overall viewing experience and reducing eyestrain.
\end{abstract}


\acmformat{Yongtao Hu, Jan Kautz, Yizhou Yu and Wenping Wang. 2013. Speaker-following Video Subtitles.}

\begin{document}

\begin{bottomstuff}

\end{bottomstuff}

\maketitle

\section{Introduction}
Subtitles are necessary for television programs, movies and other visual media for people with hearing impairments to help them understand and follow the dialog. Translated subtitles are necessary for media in a foreign language that is not dubbed to help viewers understand the spoken dialog. Subtitling is also useful for learning foreign languages.

\begin{figure}[Hhtbp]
\centering
\includegraphics[width=0.45\textwidth]{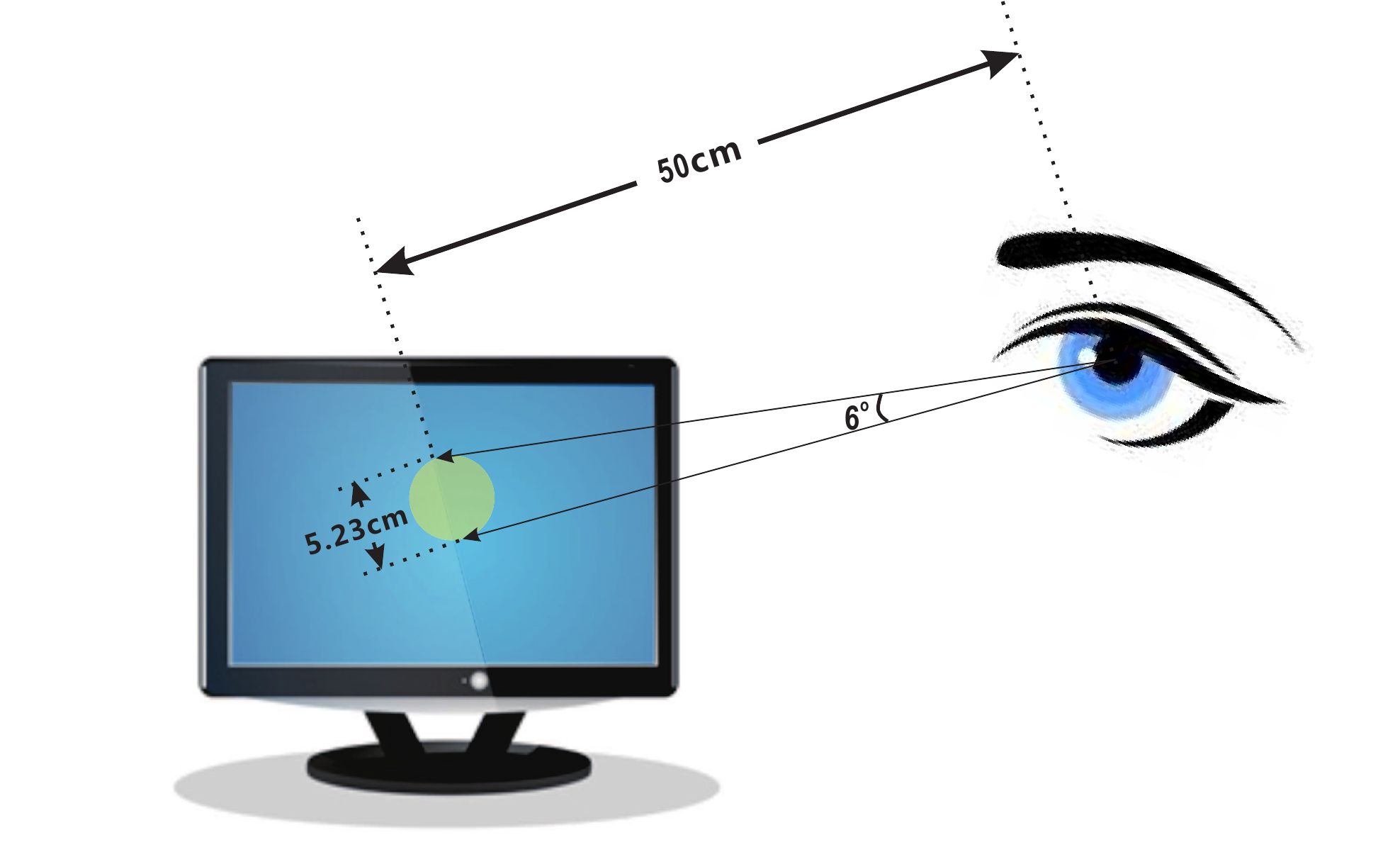}
\caption{Limited vision span of human eyes. The human vision span is an angular span of 6 degrees of arc both vertically and horizontally, which yields a region with a diameter of 5.23cm when viewed from 50cm away.}
\label{fig_vision_span}
\end{figure}

Conventionally, the position of the subtitles is in a fixed location such as at the bottom of the screen. Humans can see objects within their field of vision but can only read text clearly in a narrow vision span. This vision span is an angular span (vertically and horizontally) of approximately 6 degrees of arc, which yields a region with a diameter of 5.23cm when viewed from 50cm away (Figure \ref{fig_vision_span}) \cite{rayner1975perceptual,mcconkie1989eye,just1987psychology,wiki_vision_span}. The conventional location of subtitles at the bottom of the screen means that to follow both the facial expression of speakers and the subtitles, the viewer has to constantly move their eye gaze between the main viewing area and the bottom of the screen, leading to a high level of eyestrain.

\begin{figure}[htbp]
\centering
\subfloat[Comic word balloons \protect\cite{kurlander1996comic} \label{fig_PR_a}] {\includegraphics[height=0.19\textwidth]{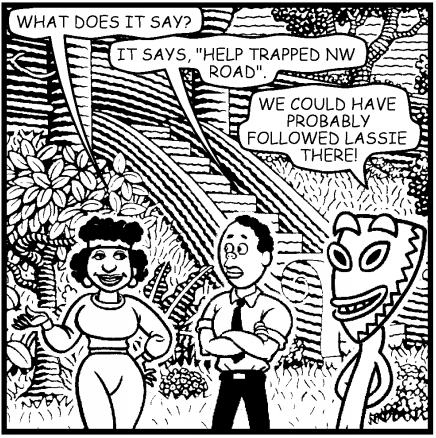}}
\hspace{0.01in}
\subfloat[Comic word balloons \protect\cite{chun2006automated} \label{fig_PR_b}] {\includegraphics[height=0.19\textwidth]{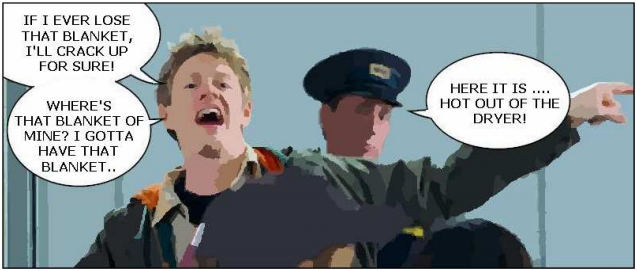}}
\hspace{0.02in}
\subfloat[Video subtitle placement \protect\cite{hong2010dynamic}]{\label{fig_PR_c} \includegraphics[height=0.19\textwidth]{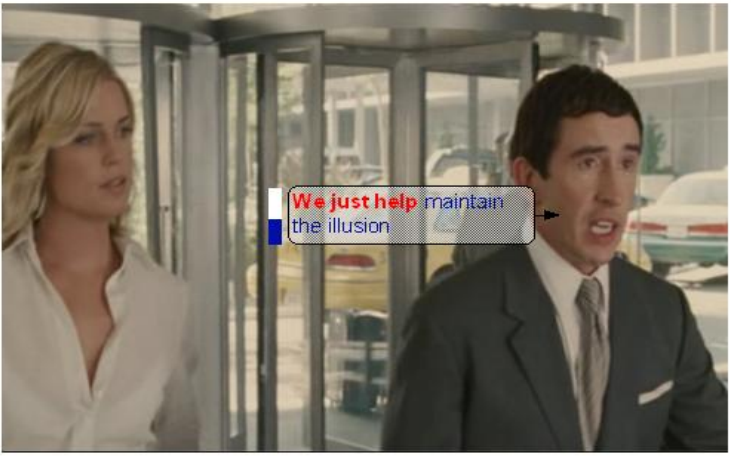}}
\caption{Previous work to improve accessibility and understanding of conversations through word balloon layout in comics (a, b), and subtitle placement in videos (c).}
\label{fig_previous_results}
\end{figure}

Considerable effort has been made to enable the viewer to understand conversations through better presentation of the spoken dialog. In comics, the word balloon construction and layout of the graphics are techniques used to achieve this goal \cite{kurlander1996comic,chun2006automated}. Word balloon layouts in comics are used to help readers better visualize the written dialog, and the layout is carefully arranged to ensure that the flow of the story is clear (Figure \ref{fig_PR_a} and \subref{fig_PR_b}). The previous work by Hong et al. \citeyear{hong2010dynamic} is the first and only research that extends these concepts to enhance the accessibility of videos by placing on screen subtitles next to the respective speaker (Figure \ref{fig_PR_c}).

\begin{figure}[htbp]
\centering
\includegraphics[width=\textwidth]{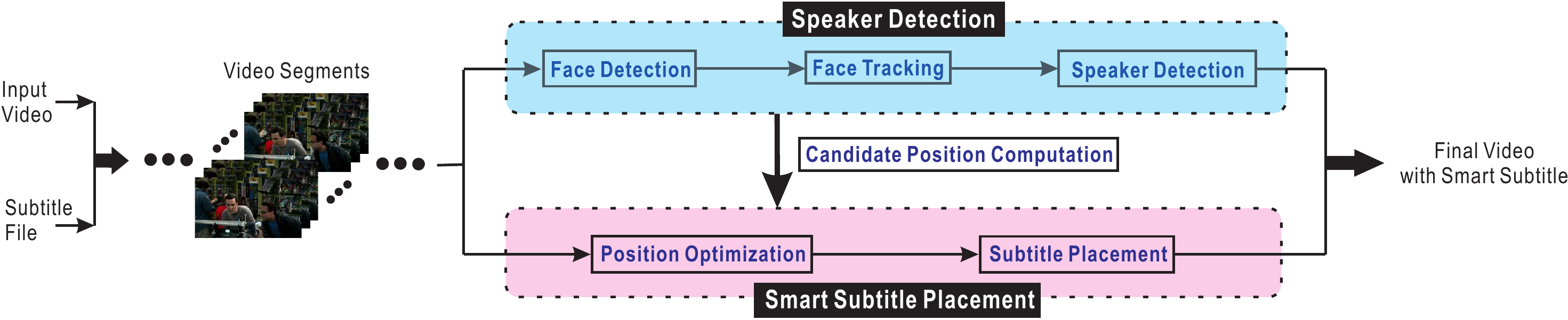}
\caption{Our framework consists of two main components: 1) speaker detection; and 2) smart subtitle placement.}
\label{fig_workflow}
\end{figure}

In this paper, we shall analyze the placement of subtitles so the viewer can comfortably follow the video content and subtitles simultaneously, which enhances the viewing experience. We shall improve on the work by Hong et al. \citeyear{hong2010dynamic} in two main aspects. We present a new speaker detection algorithm to accurately detect the speakers based on visual and audio information. An efficient optimization algorithm is then used to place the subtitle based on a number of factors. The framework of our approach is shown in Figure \ref{fig_workflow}.

To summarize, the contributions of this paper are:

\begin{itemize}
\item Development of an automatic end-to-end system for subtitle placement for TV/movies.
\item A new algorithm for effectively detecting speakers based on visual and audio information.
\item An efficient optimization algorithm for determining appropriate locations to place subtitles.
\end{itemize}

The rest of the paper is organized as follows. Related works are discussed in Section \ref{sec:related_work}. The problem formulation and the whole framework are introduced in Section \ref{sec:main_problem}. Experiments results are presented in Section~\ref{sec:experiment_results}, followed by conclusions and future works in Section \ref{sec:conclusion}.

\section{Related Work}
\label{sec:related_work}

The work by Hong et al. \citeyear{hong2010dynamic} is the first and only previous research to study how dynamically placed subtitles can enhance video accessibility for the hearing impaired. The speaker is determined by examining lip motion features and then subtitles are placed in a non-salient region (based on a saliency map) around the speaker. Their usability study showed that their method effectively improved video accessibility for viewers with hearing impairments. We have extended this work to improve video subtitling and enhance viewing experience. First, we developed a new algorithm for improving speaker detection accuracy. The method is based on both visual and audio information rather than using only lip motion because other people in the scene may also be moving their lips. Second, the subtitles are positioned using an optimization procedure that takes several factors into consideration, including absence of speakers, cross-frame coherence, and screen layout. The combined use of these factors has proven to be more robust than using image saliency alone.

Speaker detection in video is an active research topic. Speaker diarization is the problem of determining \enquote{who spoke when} based on audio signals ~\cite{anguera2012speaker}. The problem of speech recognition based on visual information has been studied  by Gordan et al. \citeyear{gordan2002support} and Saenko et al. \citeyear{saenko2004articulatory,saenko2005visual}. Visual and audio data are often used together in speaker localization (\cite{nock2003speaker,potamianos2003recent}). Generally speaking, methods based on training data are more accurate but are more time-consuming. Out of efficiency consideration, we developed a method without involving a training process but still achieved robust detection results because of our use of several novel features and cues.

Subtitle placement has been previously investigated by Kurlander et al. \citeyear{kurlander1996comic} and Chun et al. \citeyear{chun2006automated} in their work on word balloon layouts in comics. Their methods are focused on handling single frames and so cannot easily be extended to video subtitles because they cannot ensure cross-frame coherence. In contrast, our approach employs a comprehensive optimization procedure to ensure cross-frame coherence to improve video subtitle presentation.

\section{Methodology}
\label{sec:main_problem}

\subsection{Problem Formulation}

Given a video (TV, movie, etc.) and its subtitle (caption) file as input, our goal is to generate the same video but with the subtitles positioned next to their corresponding speakers to provide better viewing experience. Our method consists of two main components: 1) speaker detection; and 2) optimization of subtitle positions. The workflow of the method is shown in Figure \ref{fig_workflow}.

\subsection{Preliminaries}

\begin{figure}[htbp]
\centering
\includegraphics[width=0.6\textwidth]{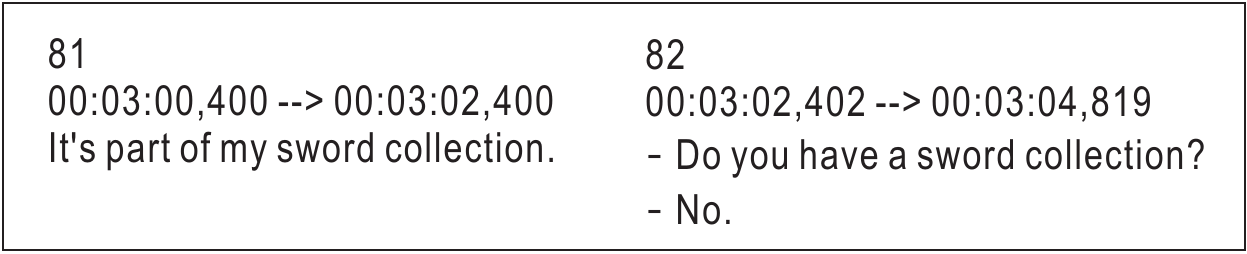}
\caption{Two examples of subtitle segments. A subtitle segment has three parts: the segment index (the first line), the timing information of the segment (the second line),  and the words spoken in this segment (the third line). The first subtitle segment is for one speaker and the second subtitle segment for two speakers.}
\label{fig_subtitle}
\end{figure}

The subtitle file is a text file in SRT format consisting of subtitle segments that contain the spoken lines and timing information. Time information provides only an approximate time interval during which the subtitle are shown on the video screen. Since this interval is always longer than the speaking time with extra time added at the beginning and the end. The input video consists of video segments corresponding the timing information in the subtitle file. Those video segments that have corresponding subtitle segments are called~\enquote{\textit{speaking video segments}}, while the others called \enquote{\textit{non-speaking video segments}}. Figure \ref{fig_subtitle} shows an example of two consecutive subtitle segments. A subtitle segment may contain lines from more than one speaker, as in the second example. The first example in Figure \ref{fig_subtitle} corresponds to one speaking video segment, while the second example to two speaking video segments for two speakers.

We shall explain next how to process all the speaking video segments for speaker detection and optimization of subtitle placement.

\subsection{Speaker Detection}
For each speaking video segment, we perform the speaker detection procedure to detect the speakers. The workflow for speaker detection is shown in light blue in Figure \ref{fig_workflow}. Our strategy is to initially detect the faces and combine them to form face tracklets. We examine all the face tracklets to determine the speaker for a given subtitle segment. The details of each step are discussed below.

\subsubsection{Face Tracking}
\label{subsubsec:face_tracking}

We first detect faces using the OpenCV frontal + profile Viola-Jones face detector \cite{viola2004robust}. Then these detected faces are linked using a low-level association approach to obtain face tracklets, following Kuo et al. \citeyear{kuo2010multi}. Regions with similar positions, sizes and appearances are linked. These low level tracklets are further linked into final face tracklets if their size, appearance and coherent direction of movement are very similar to each other.

\begin{figure}[htbp]
\centering
\includegraphics[width=0.4\textwidth]{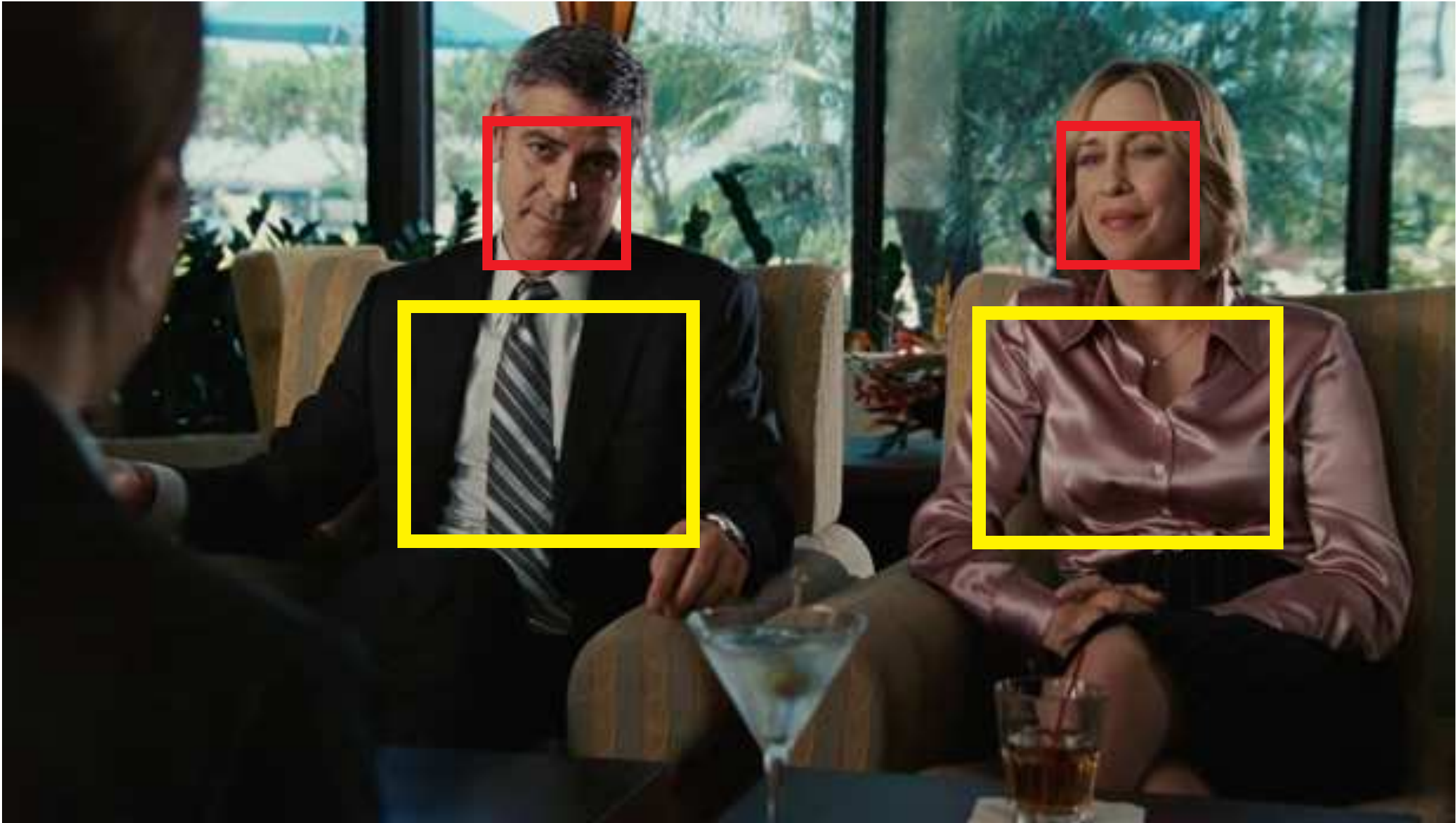}
\caption{Clothing appearance used in face tracking. Red rectangles show the detected faces and yellow rectangles show the corresponding clothing appearance.}
\label{fig3}
\end{figure}

Face detections are not directly used for face tracking because matching the appearances of faces is extremely challenging due to similar colors, expressions, poses of different faces, in addition to lighting changes and motion blur \cite{zhang2003automated}. Therefore, the clothing appearance has been used as additional cues for matching the same face \cite{jaffre2004costume}. Previous works show that face tracking in certain conditions, where matching is not possible based on face appearance alone, can be overcome by matching clothing appearances \cite{everingham2006hello}. This improvement is mainly due to the richer texture variety of clothing compared to faces. Hence, we incorporated clothing appearance in our face tracking algorithm and found that this improved the performance significantly. The bounding boxes of clothing are extracted according to the localization and the scale of the detected faces. We estimate the clothing appearance in an area under the face at a distance of $0.2 *$ face height. The size of this rectangular clothing area is proportional to the face area with a width $2d$ and height $1.5d$, where $d$ is the width of the face. Examples are given in Figure \ref{fig3}. We chose these coefficients from the best fitting of the box in  learning images.

\subsubsection{Speaker Detection}
\label{subsubsec:speaker_detection}

Lip motion is the most widely used feature for speaker detection \cite{everingham2006hello,hong2010dynamic}. However, the lip motion feature alone is not robust enough for speaker detection because the detection of a mouth region is often not accurate. We have observed that in TV and movies the positions and duration of the speakers are highly distinguishable from non-speakers. In addition to lip motion, we introduce a center contribution feature and length consistency feature for better speaker detection. We further consider the audio-visual synchrony feature to take into account the close relationship between audio and visual information to improve the robustness of our method. Based on these observation, in the presence of multiple speaker candidates, we detect the true speaker by examining the following four features: 1) {\em lip motion}, 2) {\em center contribution}, 3) {\em length consistency} and 4) {\em audio-visual synchrony}. The detailed procedures are explained below.

\paragraph{Lip Motion}
A speaker often has more significant lip motion than non-speakers. Similar to Everingham et al. \citeyear{everingham2006hello}, we can detect the speaker from speaker candidates by detecting those faces with significant lip motion. For each face tracklet, we first use a facial landmark detector \cite{uricar2012detector} to detect mouth corners. The mouth region can then easily be predicted based on the mouth corners. The mean squared difference (MSD) of the pixel values in mouth region is then computed between the current and previous frames. We compute the average of MSDs as the lip motion feature as
\begin{equation}
\mathcal{MSD} = \frac{1}{N-1} \left[\sum_{i=1}^{N-1}{\mathcal{MSD}_{(M_i, \; M_{i+1})}} \right],
\label{equ_MSD}
\end{equation}
where $N$ is the length (i.e. \# of frames) of the speaker tracklets, $M_i$ the mouth region of the speaker in the $i$-th frame.

\paragraph{Center Contribution}
It is observed that, in most TV and movies, a speaker is more likely than non-speakers to be located towards the center of the screen. To leverage this observation, we introduce the {\em center contribution} (CC) feature as
\begin{equation}
\begin{aligned}
\mathcal{CC}= \frac{1}{N}\left[ \sum_{i=1}^{N}{\mathcal{CC}_i} \right]
\text{, with } \mathcal{CC}_i= 100 \times \left[1- \frac{d\left(P_i, \; P_c\right)}{d\left(\mathcal{O}, \; P_c\right)} \right]
\end{aligned},
\label{equ_CC}
\end{equation}
where $P_i$ is the center of the speaker's face in $i$-th frame, and $\mathcal{O}$ and $P_c$  are the origin and center of the image plane, respectively. Here  $d(\cdot, \; \cdot)$ is the distance function between two points in Euclidean space.

\paragraph{Length Consistency}
The accuracy of speaker detection can be further improved by considering the consistence between the length of a candidate face tracklet and the length of speaking time of a speaking subtitle segment. We call this {\em length consistency}. Similar to MSD and CC, the candidate face tracklet with a higher LC is more likely to be the true speaker. We compute the length consistency (LC) as follows
\begin{equation}
\begin{aligned}
& \mathcal{LC} = \frac{1}{|L-L_{std}|},
& \mbox{ with} ~~ ~~
    \left\{ \begin{aligned}
        & L_{std} = \frac{L_{words} \cdot F_{video}}{\overline{V_{speaking}}}, \\
        & \overline{V_{speaking}} = \frac{L_{total\_words}}{T}, \\
    \end{aligned} \right.\\
\end{aligned}
\label{equ_LC}
\end{equation}
where $L$ is the length of the candidate face tracklet, $L_{words}$ the number of the spoken words in the current subtitle segment, and $L_{total\_words}$ the total number of words in the input video, $T$ the total speaking time in the input video, $F_{video}$ the frame rate of input video, and $\overline{V_{speaking}}$ the average speaking speed in the input video (i.e. the number of words spoken in unit time).

\paragraph{Audio-Visual Synchrony}
Previous work \cite{driver1996enhancement,wallace2004unifying} has demonstrated that audio cues and video cues can be combined to enhance the understanding of an environment. For example, sounds appear to correspond to motion synchronous with acoustic stimuli. Audio-visual (AV) synchrony has been used to resolve speaker localization \cite{monaci2011towards} by computing the synchronization score of audio with the lower half of the faces, with audio-visual co-occurrence measured by the \textit{synchronization score} $\mathcal{AV}$ of time slot $\Delta=[T_1, T_2]$ which is computed as
\begin{equation}
\mathcal{AV} = \left< y_a(t), \; y_v(t) \right>, \text{ with } t \in \Delta,
\label{equ_AV}
\end{equation}
where $y_a, \; y_v$ are the audio and visual feature vectors and $\left< \cdot, \cdot \right>$ indicates the scalar product between the vectors.

\begin{figure}[htbp]
\centering
\includegraphics[width=0.8\textwidth]{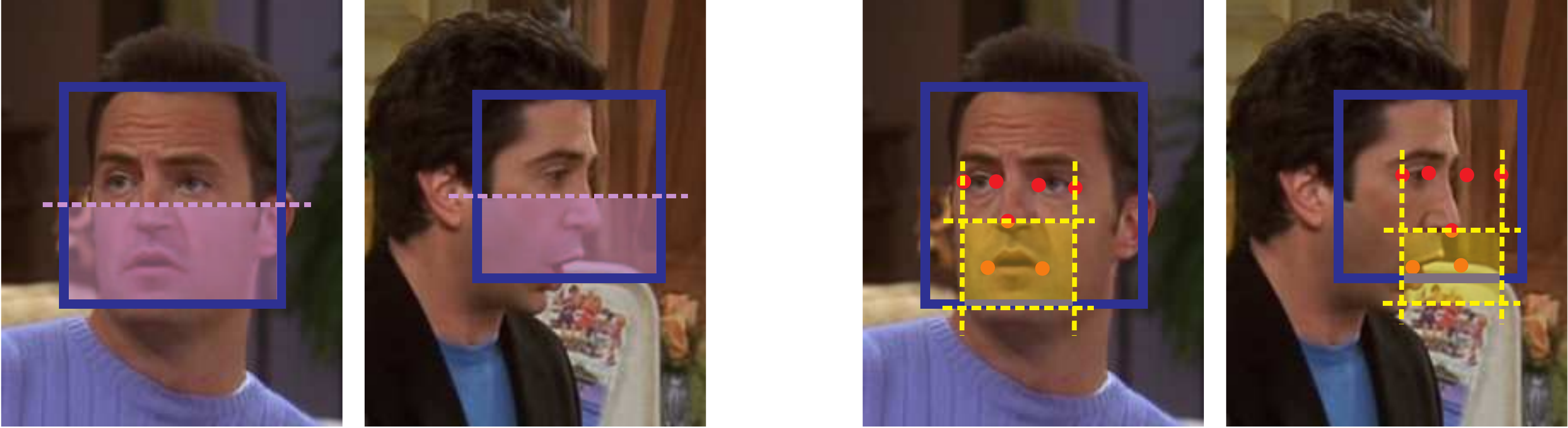}
\caption{Improving audio-visual synchrony through better motion region prediction. Left two: motion region (light pink area) used by Monaci \protect\citeyear{monaci2011towards}. Right two: better motion region (light yellow area) prediction based on facial landmarks (red points) from our method.}
\label{fig_AV_compare}
\end{figure}

Speaker detection that is solely based on lip motion \cite{hong2010dynamic} relies too much on the accuracy of the mouth region detection and therefore may fail for complex scenes. Our method includes {\em center contribution} and {\em length consistency} features, and we further improve speaker detection by examining {\em audio-visual synchrony}. Moreover, unlike Monaci \citeyear{monaci2011towards}, we use extended facial landmarks \cite{uricar2012detector} for better prediction of the motion region instead of using only the lower half of the face when examining audio-visual synchrony. This will help to filter out any motion disturbance in the background, especially when speakers or cameras are moving. Our method also compensates for the errors due to poor face detection and facial landmarks (see Figure \ref{fig_AV_compare}).

\begin{figure}[htbp]
\centering
\includegraphics[width=0.8\textwidth]{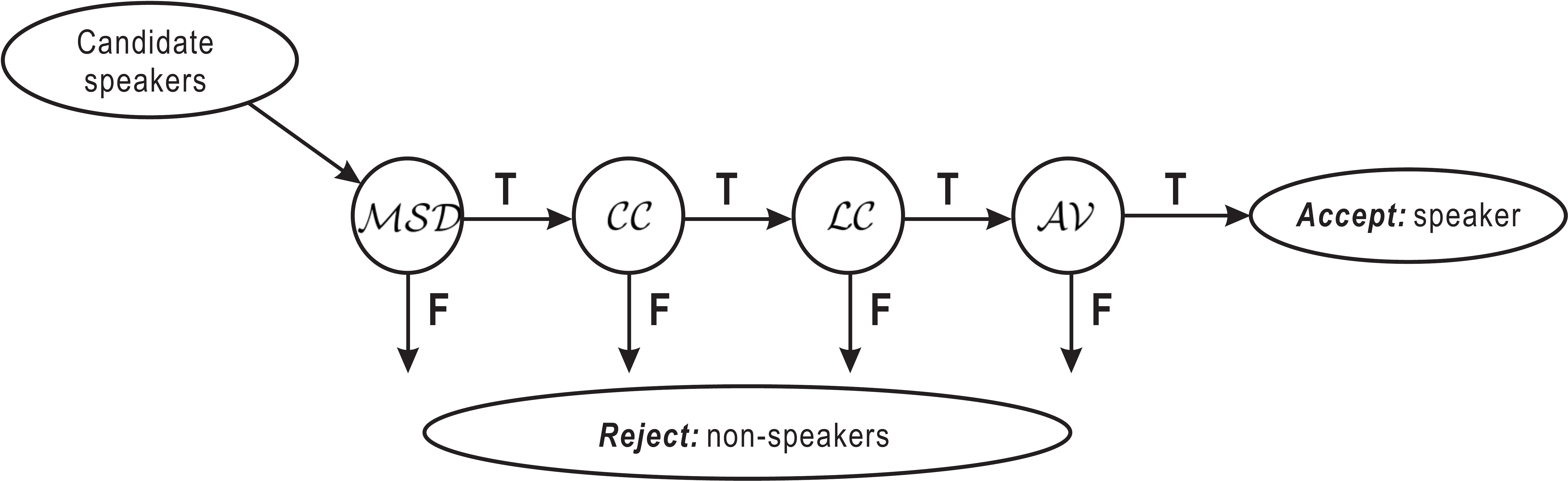}
\caption{Schematic depiction of our cascade speaker detection method. Only candidate speakers that pass the current test  will go through to the next level.}
\label{fig_cascade_speaker_identification}
\end{figure}

For better accuracy and efficiency, we chain the four features above in a cascade in the following order: {\em lip motion}, {\em center contribution}, {\em length consistency} and {\em audio-visual synchrony} (see Figure \ref{fig_cascade_speaker_identification}).  The process can be seen as a degenerate decision tree \cite{viola2004robust}. The design of this cascade structure is based on the observation that the majority of the candidate speakers are non-speakers and only candidate speakers that satisfy the current constraint pass to the next level. We put {\em lip motion} first in the cascade because it can effectively reject most non-speakers. {\em Center contribution} and {\em length consistency} are put second and third in the cascade as they can be evaluated very quickly. {\em Audio-visual synchrony} is placed last in the chain because this evaluation is more time-consuming. Experiments have also shown that a cascade in this order gives the best results. Detailed implementation is shown in Algorithm \ref{algor_speaker_detection}\footnote{We apply $\theta_1=20$, $\theta_2=2.5$, $\theta_3=2$, $\theta_4=0.1$, $\theta_5=2$ throughout.}.

Figure \ref{fig_draw} shows the performance improvement in speaker detection accuracy\footnote{Speaker detection accuracy is computed in terms of video segments. It is considered to be correct for those video segments where the subtitle is put in the default position (i.e. the bottom of the screen) when no speaker is detected.}. The left figure shows the effect of adding the {\em center contribution}, {\em length consistency} and {\em audio-visual synchrony} (with better motion prediction) besides {\em lip motion}, and the right figure shows our speaker detection accuracy compared with previous methods \cite{monaci2011towards,hong2010dynamic}\footnote{Hong et al. \citeyear{hong2010dynamic} used only the {\em lip motion} feature (i.e. MSD) for speaker detection.}. Overall, our speaker detection method outperformed these two methods.

\begin{algorithm}[Hhtbp]
\caption{Speaker detection algorithm.}
\label{algor_speaker_detection}
\SetKwInOut{Input}{Input}
\SetKwInOut{Output}{Output}
\Input{$N$ face tracklets, each of which stands for a speaker candidate.}
\Output{The speaker (tracklet) or no speaker.}
\BlankLine
    Compute $\mathcal{MSD}$ feature for each face tracklet. \\
    Delete all tracklets whose $\mathcal{MSD}$ $< \theta_1$ (assume remaining $N_1$ tracklets after this). \\
    \uIf{$N_1 >= 2$}
    {
        Delete all tracklets whose $\mathcal{MSD}$ $* \theta_2 < \max_{\mathcal{MSD}}$ (assume remaining $N_2$ tracklets after this). \\
        \uIf{$N_2 >= 2$}
        {
            Compute $\mathcal{CC}$ feature for each face tracklet. \\
            Delete all tracklets whose $\mathcal{CC}$ $* \theta_3 < \max_{\mathcal{CC}}$ (assume remaining $N_3$ tracklets after this). \\
            \uIf{$N_3 >= 2$}
            {
                Compute $\mathcal{LC}$ feature for each face tracklet. \\
                \lIf{$\max_{\mathcal{LC}} - second \max_{\mathcal{LC}} > \theta_4$}
                {
                    \Return the tracklet with $\max_{\mathcal{LC}}$. \newline
                }
                \Else
                {
                    Compute motion region based on facial landmarks for each tracklet.\\
                    Compute $\mathcal{AV}$ feature for each face tracklet \\
                    \lIf{$\max_{\mathcal{AV}} > \theta_5$}
                    {
                        \Return the tracklet with $\max_{\mathcal{AV}}$. \newline
                    }
                    \lElse
                    {
                        \Return no speaker.
                    }
                }
            }
            \lElseIf{$N_3 == 1$}
            {
                \Return current tracklet.
            }
            \lElse
            {
                \Return no speaker.
            }
        }
        \lElseIf{$N_2 == 1$}
        {
            \Return current tracklet.
        }
        \lElse
        {
            \Return no speaker.
        }
    }
    \lElseIf{$N_1 == 1$}
    {
        \Return current tracklet.
    }
    \lElse
    {
        \Return no speaker.
    }
\BlankLine
\end{algorithm}

\subsection{Subtitle Placement}
Generally speaking, our goal is to place subtitles close to their corresponding speakers. There are several considerations in attaining this goal: 1) In each frame, the subtitle should be close enough to its speaker so not to cause any confusion that it is spoken by another person; 2) Across consecutive frames, the overall distance between all subtitle placements should be small to reduce eyestrain; 3) Subtitles should not be placed near screen boundaries so not to detract the viewer from the central viewing area; 4)  Subtitles should not occlude any important visual contents (e.g. faces).

In this section, we shall present an optimization framework (detailed in in Algorithm \ref{algor1}) for subtitle placement. First, we shall explain some preprocessing steps before performing subtitle placement.  The following operations are needed to process the input video file and the subtitle file.

\begin{enumerate}
\item Split a speaking video segment when the speaker moves significantly during a subtitle segment;
\item Split a subtitle segment if there is a major shot change in the corresponding video segment;
\item Split a subtitle segment when it corresponds to multiple speakers; and
\item Refine the display time of a subtitle from the provided timing information.
\end{enumerate}

\begin{algorithm}[Hhtbp]
\caption{Subtitle placement algorithm.}
\label{algor1}
\SetKwInOut{Input}{Input}
\SetKwInOut{Output}{Output}
\Input{Speaker detection result (i.e. face tracklets for each subtitle segment).}
\Output{Positions for subtitle placement at each subtitle segment.}
\BlankLine
\ForEach{subtitle segment}
{
    Determine the proper rectangle to hold the subtitle\footnotemark.\\
    Compute candidate positions for placing the subtitle.
}
Compute the optimal positions of subtitle based on candidate positions. \\
\BlankLine
\Return the position of each subtitle segment.
\end{algorithm}

\footnotetext{Similar to Hong et al. \citeyear{hong2010dynamic}, we use a rectangle to bound the subtitle and an arrow to the speaker to avoid confusion. The dimensions of the bounding rectangle of the subtitle can easily be determined by processing the subtitle text based on its length and font.}

\subsubsection{Computing Candidate Subtitle Positions}
\begin{figure}[Hhtbp]
\centering
\includegraphics[width=0.4\textwidth]{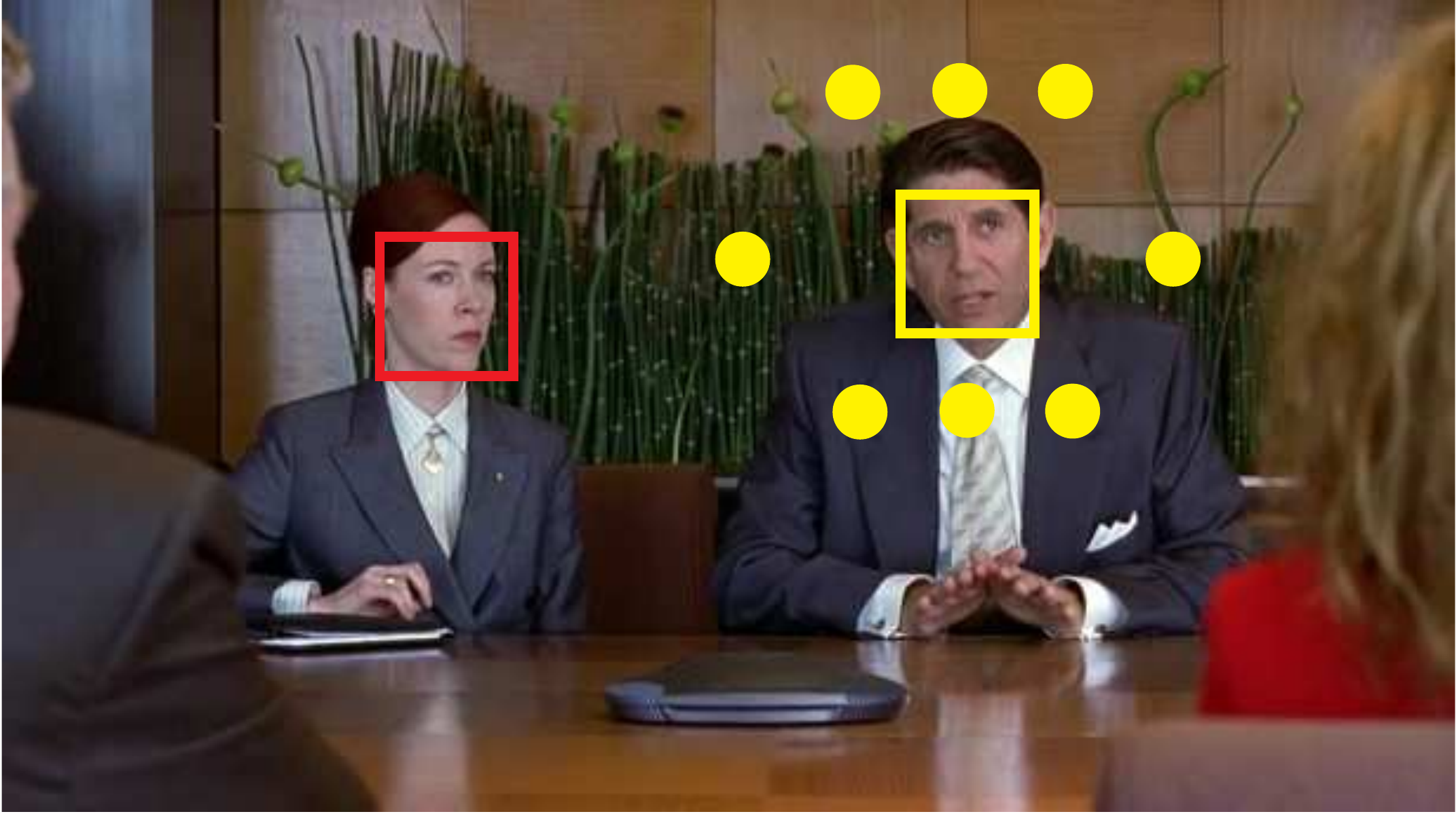}
\caption{Candidate positions for the placement of the subtitle around the speaker's face. Speaker (yellow rectangle), non-speaker (red rectangle), candidate subtitle positions (yellow circles, 8 positions in our experiment).}
\label{fig_candidate}
\end{figure}

Any position close to the speaker may be a good subtitle position. In order to reduce the search space for better efficiency, we only consider the following eight candidate positions around the speaker's face: above left, above, above right, below left, below, below right, left and right, which are indicated by the yellow dots in Figure \ref{fig_candidate}). Note that the exact locations of these eight candidate positions are computed based on the speaker's location and size of the speaker's face and the length and font size of the subtitle.

\subsubsection{Subtitle Position Optimization}
Subtitle position optimization is the key step in Algorithm \ref{algor1}. We take into account layout information to obtain the best position to place the subtitle. Here, we assume that each subtitle will be assigned a fixed position.

\paragraph{Local optimization}
Locally, we require the subtitle to be close to the speaker and far away from non-speakers. Let $P$ denote the optimal subtitle position of the current subtitle segment, $P_S$ the speaker's position, and $P_{NS_k} (k=1, 2,..., K)$ the positions of the $K$ non-speakers in the scene. The energy function of $P$ due to local optimization is defined as
\begin{equation}
E_{local} = d_{\left(P, \; P_S\right)} - \sum_{k=1}^K{d_{\left(P, \; P_{NS_k}\right)}}
\label{eqa_local}
\end{equation}

\paragraph{Position consistency over time}
We require that the positions of consecutive subtitles do not change their positions too much across consecutive frames in order to alleviate eye-strain. To this end, we constrain subtitles of subsequent speaking video segments to follow the positions of the preceding subtitles. Let $P_{pre\_opt}$ denote the optimaized position of the preceding subtitle. Then the energy function due to this consideration is defined as
\begin{equation}
E_{global} = d_{\left(P, \; P_{pre\_opt}\right)}
\label{eqa_global}
\end{equation}

\paragraph{Layout with respect to screen boundary}
Subtitles should not be placed at the screen boundaries in order to allow the viewer to focus more on the central viewing part of the screen. This preference is reflected in following energy function:
\begin{equation}
E_{layout} = d_{\left(P, \; boundary\right)}
\end{equation}

Now, combining these three energy terms, we defined the following total energy function.
\begin{equation}
\min \quad E = w_1 \cdot E_{local} + w_2 \cdot E_{global} + w_3 \cdot E_{layout}.
\end{equation}
We use $w_1=1.0$, $w_2=1.0$, $w_3=-1.0$ for all our experiments.  We minimize this function to compute the optimized position of the current subtitle.

\subsection{Some Further Considerations}

\subsubsection{Advanced Video Segment Splitting}
When the speaker moves around during a subtitle segment, we would still like to have the subtitle tagged to the speaker. However, rather than letting the subtitle float with the speaker, we split both the subtitle segment and its corresponding speaking video segment into shorter segments. Such subtitle segments are first identified, based on the speaker tracklet, and are split into shorter subtitle segments. Then the positions of these shorter subtitle segments are determined using the same procedure (Algorithm \ref{algor1}) as for the normal subtitles, so that the viewer can better follow the moving speaker together with the subtitles. Note that, the position of each short subtitle is fixed for each small video segment.  One example is shown in Figure \ref{fig_result}a.

\subsubsection{Shot Change Handling}
\begin{figure}[Hhtbp]
\centering
\includegraphics[width=0.9\textwidth]{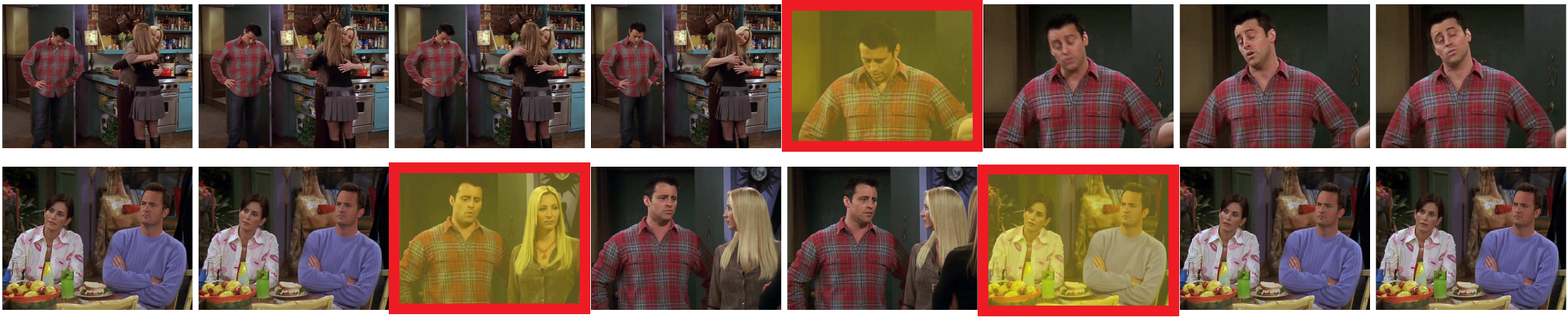}
\caption{Results of shot change detections. Each row shows a video segment. The first frame of the shot change is marked by a red rectangle with a yellow shade. There is one shot change in the first video segment and two shot changes in the second segment.}
\label{fig_scene_change}
\end{figure}

A significant shot change in TV/movies may make a speaker appear or disappear. In this case, subtitle positions need to change accordingly. For example, it should be placed at a default position (such as the bottom of the screen) if the speaker disappears after a shot change. We propose a simple shot change detector for video segments solely based on color histogram (see Algorithm \ref{algor_scene_change}\footnote{We computed the correlation coefficient as the similarity of their RGB color histogram and applied a threshold of 0.99.}) instead of using shot change detectors based on complex models \cite{sethi1995statistical,faernando2001scene,dimou2005scene}. Our shot change detector proves to be accurate enough even for some very challenging TV series such as \enquote{Friends} and \enquote{The.Big.Bang.Theory}, as shown in Table \ref{tab_scene_change}. Some results of detected shot changes are shown in Figure \ref{fig_scene_change}.

\begin{algorithm}[Hhtbp]
\caption{Shot change detector for video segments.}
\label{algor_scene_change}
\SetKwInOut{Input}{Input}
\SetKwInOut{Output}{Output}
\Input{A video segment.}
\Output{All shot change positions.}
\BlankLine
\ForEach{pair of adjacent two frames}
{
    Compute the similarity of their color histogram as $\delta$. \\
    \If{$\delta < threshold$}
    {
        Label the second frame as a shot change position.
    }
}
\BlankLine
\Return all these labeled positions.
\end{algorithm}

\begin{table}[htbp]
\centering
\begin{tabular}{c||c|c|c|c|c|c}
\hline
{\bf Video Input} & {\bf True} & {\bf Detected} & {\bf False} & {\bf Missed} & {\bf Recall} & {\bf Precision} \\
\hline \hline
Friends.S01E05 & 115 & 105 & 3 & 13 & 88.70\% & 97.14\% \\
\hline
Friends.S01E13 & 103 & 96 & 2 & 9 & 91.26\% & 97.92\% \\
\hline
Friends.S10E15 & 106 & 108 & 3 & 1 & 99.06\% & 97.22\% \\
\hline
The.Big.Bang.Theory.S03E05 & 31 & 32 & 1 & 0 & 100\% & 96.88\% \\
\hline
\end{tabular}
\caption{Performance of our shot change detection procedure. The ground truth of shot changes are manually labeled in our experiments.}
\label{tab_scene_change}
\end{table}

Based on the shot change detection, video segments are further split into shorter video segments, each of which has no big significant shot changes and the subtitle position in it can be kept relatively constant. After splitting, the video segment containing the scene that has the longest time overlap with the speaking video subtitle will be assigned the subtitle. The subtitle for all the other video segments will be set at a default position (e.g. the bottom of the screen). Some sample outputs are shown in Figure \ref{fig_result}b.

\subsubsection{Splitting Multiple Speakers}
There are often two or more speakers in a single subtitle segment, such as in the second example in Figure \ref{fig_subtitle}, where different speakers are denoted by a \enquote*{-} in these segments. In this case, we split such a subtitle segment into different parts corresponding to different speakers, using the separating character \enquote*{-}. Accurate active speaking time for each speaker is determined based on all the face tracklets of the whole segment. Then, within each small time range, Algorithm \ref{algor_speaker_detection} is used to find the specific speaker. One example is shown in Figure \ref{fig_result}c.

\subsubsection{Refining the Speaking Time}
The time information in the subtitle file is usually not the actual speaking time but only provides an approximate interval to show the subtitle on screen. The time interval is manually specified and is normally longer than the actual speaking time, with extra time padded at the beginning and the end. We refine the speaking time based on the span of the speaker tracklet, which is more reliable and coherent with the video content. Some examples are shown in Figures \ref{fig_result}a, b and c.

\section{Experiment Results and Discussion}
\label{sec:experiment_results}

\begin{figure}[htbp]
\centering
\includegraphics[width=0.95\textwidth]{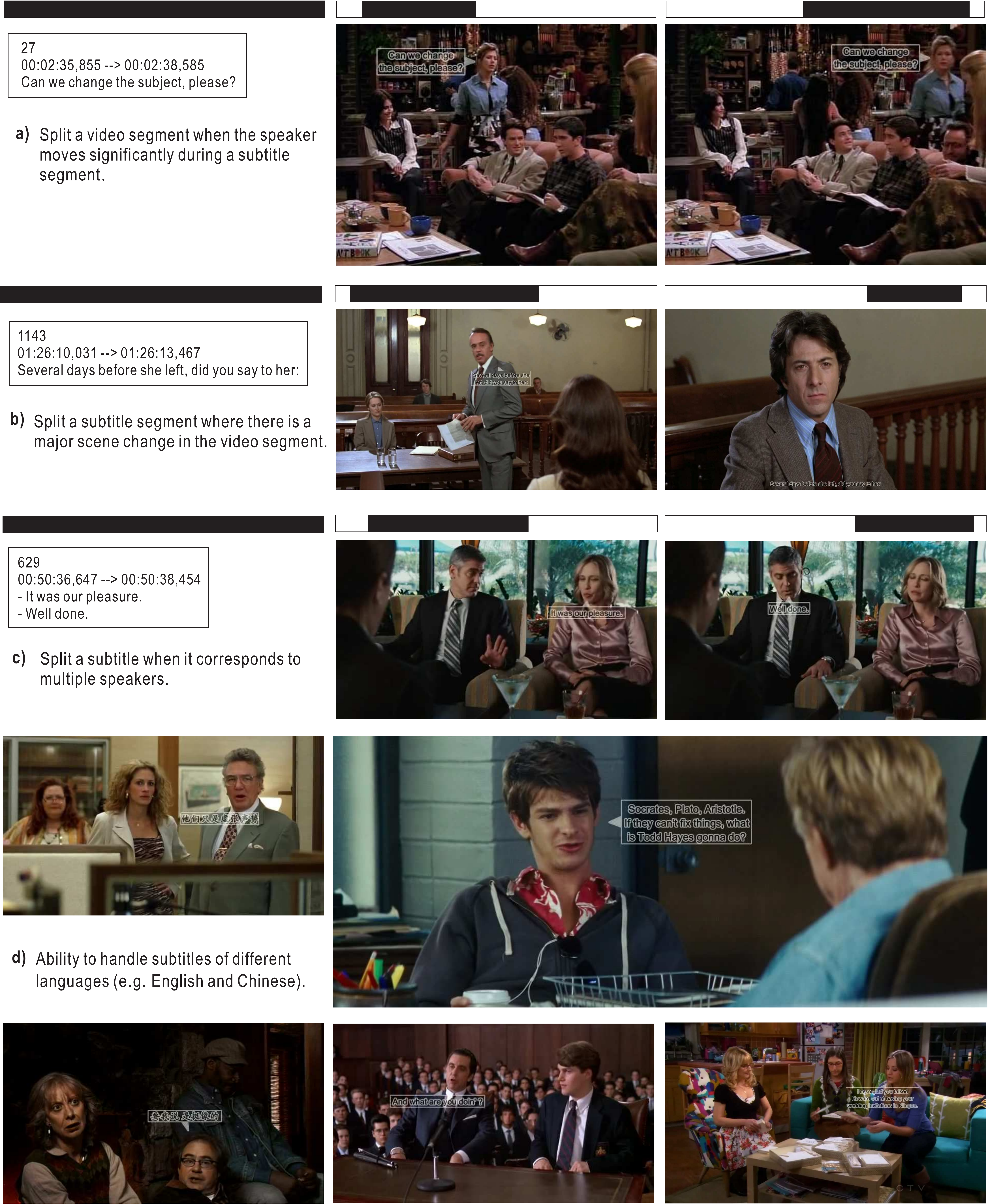}
\caption{Sample results of our system. In a) b) c), black/white bars are the active time (active in black, non-active in white).}
\label{fig_result}
\end{figure}

We tested our system on a variety of videos including those in which people speak very fast and move quickly. The videos were from the following TV series and movies: \enquote{{\em Friends}}, \enquote{{\em The Big Bang Theory}}, \enquote{{\em Kramer vs. Kramer}}, \enquote{{\em Erin Brockovich}}, \enquote{{\em Up in the Air}}, \enquote{{\em The Man From Earth}}, \enquote{{\em Scent of a Woman}} and \enquote{{\em Lions for Lambs}}. All these video clips can be accessed at the website\footnote{\url{https://sites.google.com/site/smartsubtitles/}}. Some sample outputs (TV/movie screenshots) are shown in Figure \ref{fig_result}.

\subsection{Speaker Detection Accuracy}

\begin{figure}[Hhtbp]
\centering
\includegraphics[width=0.49\textwidth]{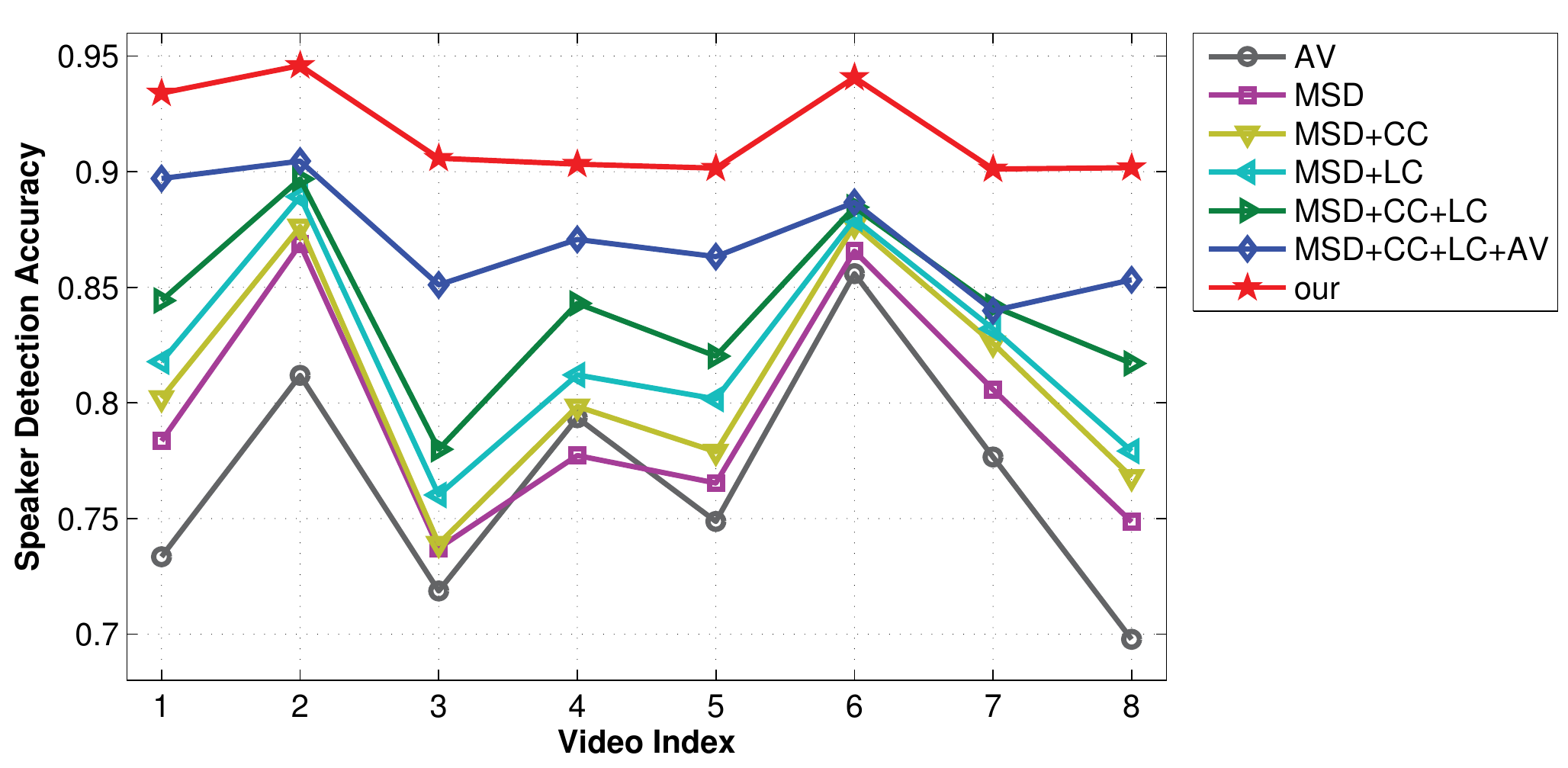}
\includegraphics[width=0.49\textwidth]{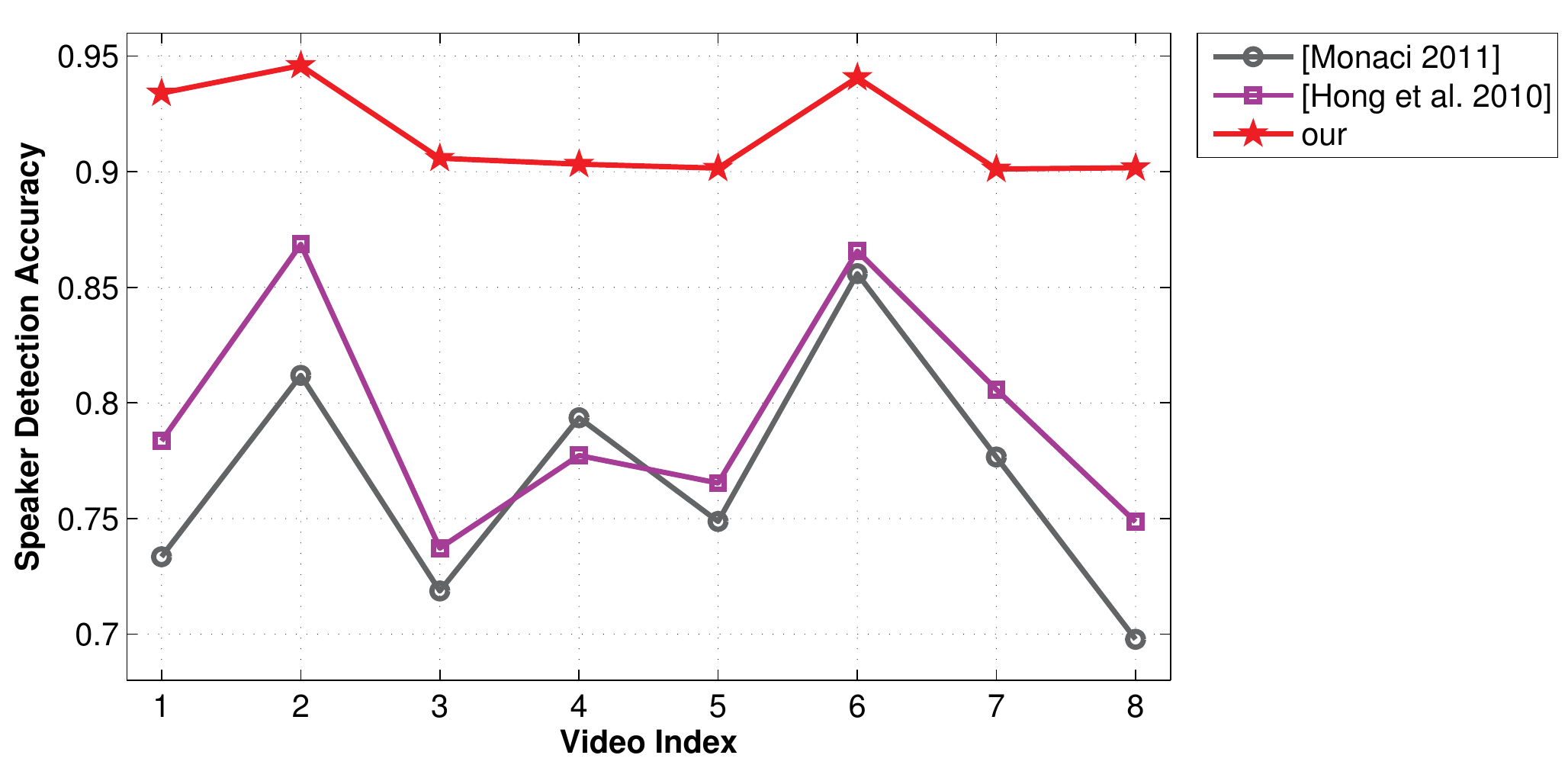}
\caption{Performance comparison in terms of speaker detection accuracy. Left: performance improvement of different features. Right: performance improvement compared with the previous method. Video indexes 1 to 8 refer to the TV clips from \enquote{Friends.S10E15} and \enquote{The.Big.Bang.Theory.S05E16}, and the movie clips from \enquote{Kramer.vs.Kramer},  \enquote{Erin.Brockovich}, \enquote{Up.In.The.Air}, \enquote{The.Man.From.Earth}, \enquote{Scent.of.a.Woman} and \enquote{Lions.for.Lambs} respectively. Ground truth speakers are manually labeled in our experiments.}
\label{fig_draw}
\end{figure}

We have introduced four new features for speaker detection. Their contributions to improving detection accuracy are shown in Figure \ref{fig_draw}. The {\em center contribution} improved the speaker detection accuracy by 0.5\%--2.5\%, {\em length consistency} improved accuracy by 1.5\%--4\%, and {\em audio-video synchrony} with better motion prediction improved accuracy by 3.5\%--6.5\%. Taken together, our speaker detection method outperformed Monaci \citeyear{monaci2011towards} by 8.5\%--20.5\% and Hong et al. \citeyear{hong2010dynamic} by 7.5\%--17\%.

\subsection{Subtitle Placement}
\begin{figure}[htbp]
\centering
\includegraphics[width=0.9\textwidth]{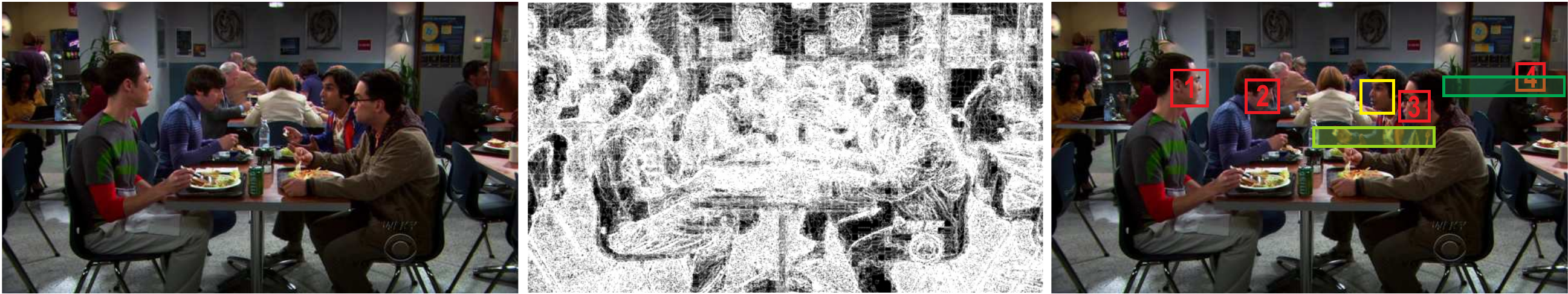}
\caption{Use of saliency map in a complex background. Left: a screen shot of the TV clip \enquote{The.Big.Bang.Theory.S05E16}; Middle: a saliency map used in the method by Hong et al. \protect\citeyear{hong2010dynamic}; Right: subtitle placement by our method (yellow shadowed rectangle) vs. that by Hong et al. \protect\citeyear{hong2010dynamic} (green shadowed rectangle) for the speaker (yellow rectangle), red rectangles are non-speakers. The subtitle placement by the method of Hong et al. \protect\citeyear{hong2010dynamic} mistakes the non-speaker \#3 to be speaking and blocks non-speaker \#4, thus confusing the viewer.}
\label{fig_saliency_problem}
\end{figure}

In the previous method \cite{hong2010dynamic}, subtitles are positioned in the least salient region around the speaker based solely on a saliency map. This performs well only with a relatively simple background. For videos with a complex background such as the example shown in Figure \ref{fig_saliency_problem}, the subtitles generated that way may block some faces. In contrast, our subtitle placement procedure is more robust for complex backgrounds. The local energy term in Equation \ref{eqa_local} tends to position the subtitle close to the speaker and at the same time far away from non-speakers, thus lessening confusion and reducing eyestrain. Furthermore, our subtitle placement algorithm is consistent across frames thanks to the global energy term in Equation \ref{eqa_global}.

\subsection{User Study}
To evaluate our system, we conducted a comprehensive usability study\footnote{\url{https://sites.google.com/site/smartsubtitles/user_study}} to compare it with conventional fixed position subtitles and a previous dynamic subtitling method.

\subsubsection{Video Clips Selection}
For the usability study, we selected 11 video clips from the above eight TV/movie videos. Specifically, two clips were selected from the movies \enquote{{\em Erin Brockovich}}, \enquote{{\em Up in the Air}} and \enquote{{\em The Man from Earth}} and one clip from each of the other five videos. Video clips were selected from video segments containing several speakers with significant switching of dialog. The average length of the 11 video clips is 2.3 minutes.

For a fair comparison, the video clips were chosen such that the speaker detection accuracy of them is consistent with that of the entire TV/movie videos from which they were selected ($\pm1.84\%$ in our user study). The consistency here is only considered for the video clips selected for our results (\textit{Dynamic\_III}) instead of to all versions (\textit{Static}, \textit{Dynamic\_I} and \textit{Dynamic\_II}) which was probably not possible. Video clips of other versions were selected accordingly.

\subsubsection{Video Group Setup}
For each of the 11 video clips, we produced a group of four different versions, namely \textit{Static}, \textit{Dynamic\_I}, \textit{Dynamic\_II} and \textit{Dynamic\_III}. Details of the version for each video group are shown in Table \ref{tab_video_versions}.

\begin{table}[htbp]
\centering
\begin{tabular}{c||l}
\hline
{\bf Video Versions} & {\bf Description}  \\
\hline \hline
Static & Traditional fixed version where subtitles are always put at the bottom of the screen  \\
\hline
Dynamic\_I & \cite{hong2010dynamic} \\
\hline
Dynamic\_II & Combine our speaker detection with \cite{hong2010dynamic}'s subtitle placement \\
\hline
Dynamic\_III & Ours \\
\hline
\end{tabular}
\caption{Details of the four versions for each video group.}
\label{tab_video_versions}
\end{table}

Note that volume demonstration and subtitle highlighting are included in the method by Hong et al. \citeyear{hong2010dynamic} because their method was mainly developed for people with hearing impairment. However, their usability study showed that the effects of volume demonstration and subtitle highlighting were minor for people with hearing impairments. Therefore, these cues were not used in the \textit{Dynamic\_I} version because it may be a distraction for the users in our study group. Users were asked to score (on a scale of 1 to 10) each of the four versions for overall viewing experience and for eyestrain level. The scores are uniformly defined with a higher score corresponding to a better overall viewing experience or a less eyestrain level.

\subsubsection{Audio and Subtitle Language Setup}
Based on the assumption that subtitles help the viewer understand the videos better, we chose specific videos with such audio so that subtitling would be useful rather than redundant. We chose clips with audio in French, Spanish and Hindi depending on the participant's language background. Subtitles, where possible, were in the native language of the reviewer, that is, Chinese for native Chinese speakers and English for native English speakers (or others).

\subsubsection{Results}
\begin{figure}[htbp]
\centering
\subfloat[Gender]{\includegraphics[width=0.235\textwidth]{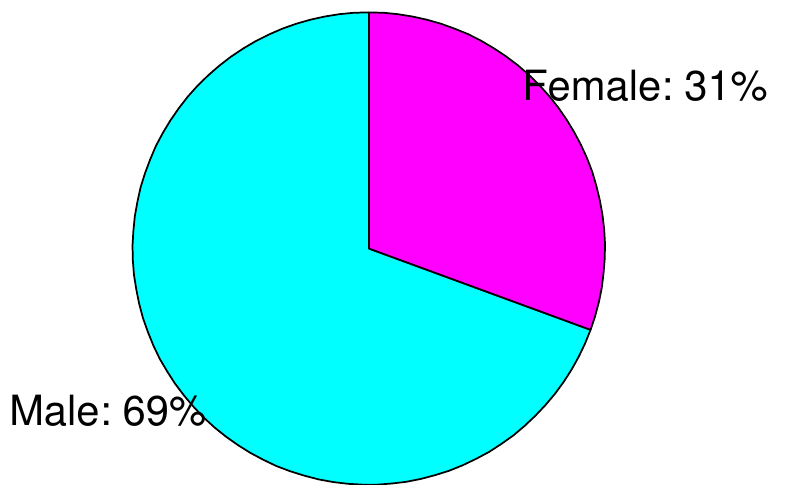}}
\subfloat[Age]{\includegraphics[width=0.255\textwidth]{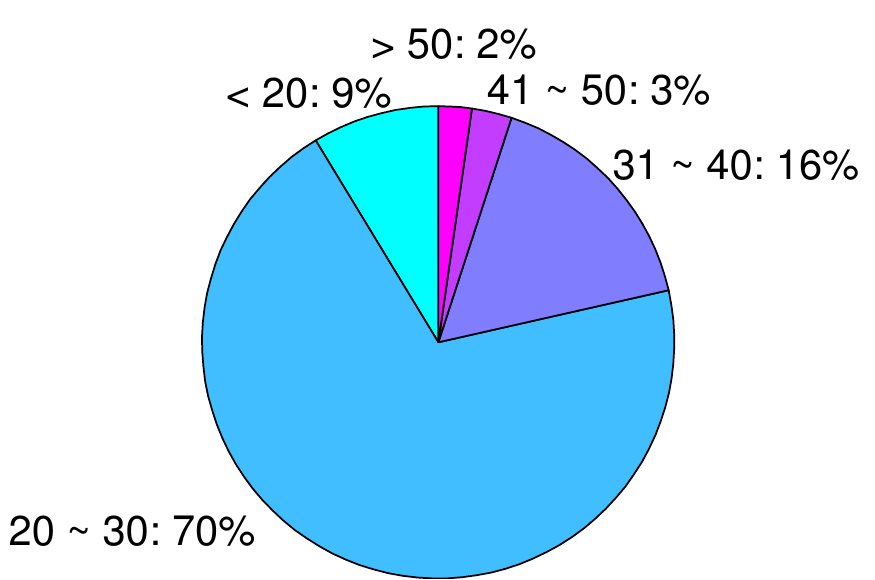}}
\subfloat[Educational background]{\includegraphics[width=0.255\textwidth]{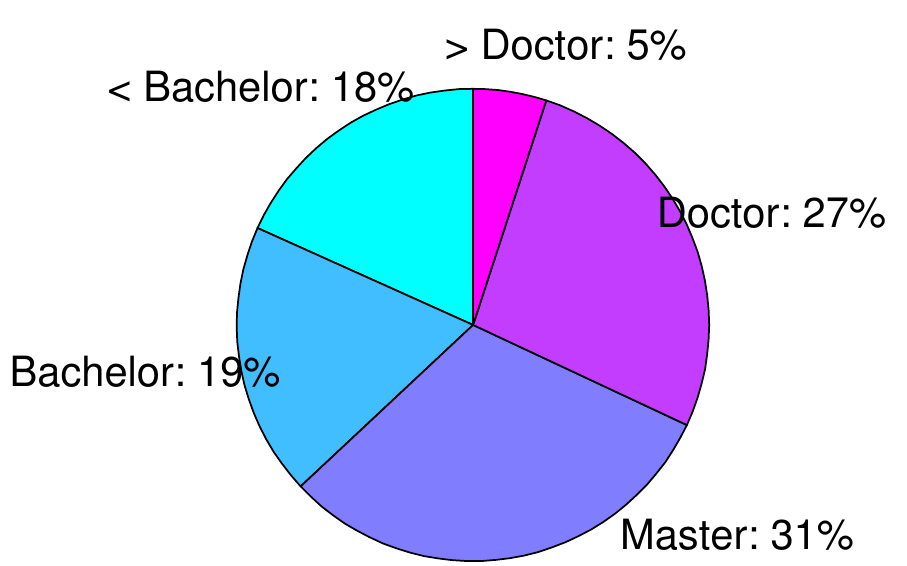}}
\subfloat[Native language]{\includegraphics[width=0.235\textwidth]{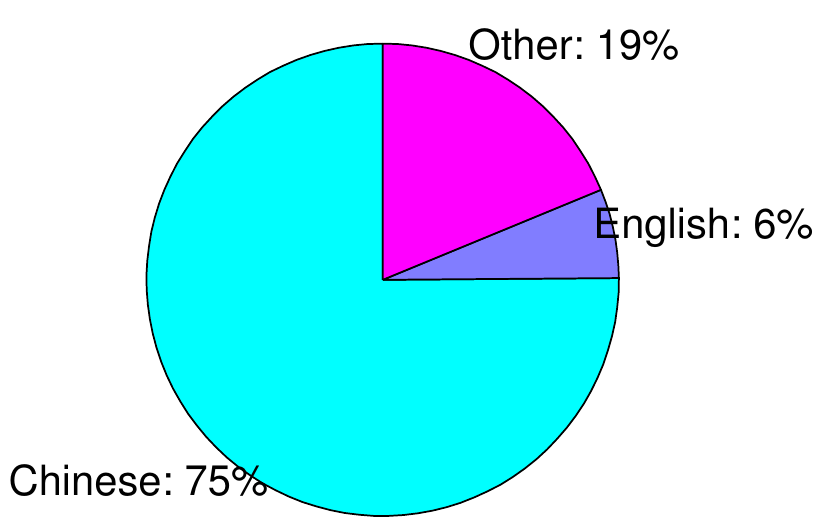}}
\caption{Information about the participants in our user study (\textit{gender}, \textit{age}, \textit{educational background} and \textit{native language}.}
\label{fig_user_study_users}
\end{figure}

Each user was asked to view at least one video group of the four versions in entirety. In total, there were 219 participants in our usability study, amounting to about 20 for each video group. The users were of various ages and genders, from many different countries, and had different language and educational backgrounds (Figure \ref{fig_user_study_users}).

\begin{figure}[htbp]
\centering
\includegraphics[width=\textwidth]{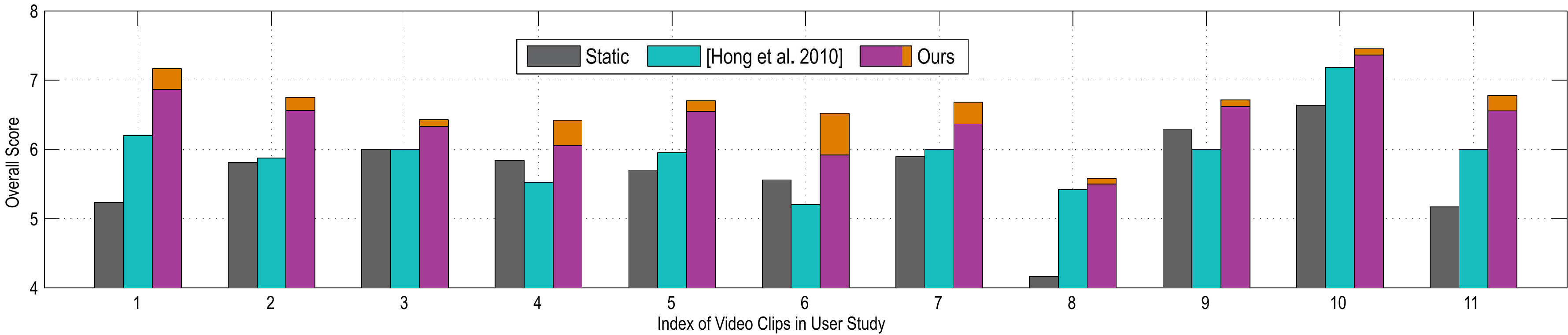}
\includegraphics[width=\textwidth]{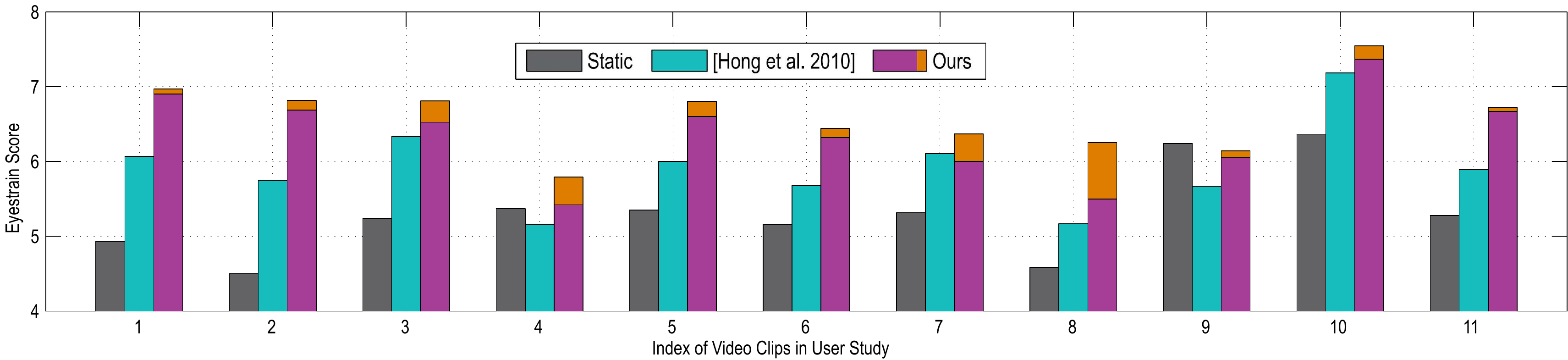}
\caption{Result of the user study. Overall viewing experience (above) and eyestrain level (below). Higher scores indicate better outcome. \textit{Dynamic\_I} version is the method by Hong et al. \protect\citeyear{hong2010dynamic}, the purple colored part of the third bar is for \textit{Dynamic\_II} version} and the whole third bar is our final version.
\label{fig_user_study_result}
\end{figure}

The results of the user study are shown in Figure \ref{fig_user_study_result}. We can see that all our results outperformed the \textit{Dynamic\_I} version \cite{hong2010dynamic} in terms of the overall viewing experience and degree of eyestrain level. In terms of overall viewing experience, the results showed that the \textit{Dynamic\_I} version \cite{hong2010dynamic} was worse or equal to the experience of Static version in 4 of the 11 groups. In comparison, the results from the \textit{Dynamic\_III} version (our method) outperformed the Static version in all groups. In terms of eyestrain, the method of Hong et al. \citeyear{hong2010dynamic} was worse than or equal to the Static version in 2 of the 11 groups, whereas our method was better than the Static version in all but one group. Nevertheless, the results for that one group (video \#9) were still better than the \textit{Dynamic\_I} version \cite{hong2010dynamic}. The \textit{Dynamic\_II} version outperformed \textit{Dynamic\_I} version in all groups for overall viewing experience and in 10 of the 11 groups for eyestrain. Our results indicate that our speaker detection is more accurate than the method by Hong et al. \citeyear{hong2010dynamic}. Moreover, our results outperformed the \textit{Dynamic\_II} versions in terms of both overall viewing experience and eyestrain level, further validating our improved subtitle presentation method.

It should be noted that, in video \#9 the level of eyestrain in all three dynamic subtitle versions were worse than in the Static version, although our method was the best. The corresponding scene in the clip showed seven people arguing with the camera shot changing rapidly, which probably caused viewers to follow the rapid switching of the subtitle position in the changing conversation.

\subsection{Applications}

Our method can be used in several applications.

\subsubsection{As a tool to assist the TV/movie industry}
The usability study has demonstrated that our system can assist the TV/movie industry as a speaker detection tool for subtitling to provide a better viewing experience. Although not 100\% accurate ($>$90\% accuracy), our speaker detection tool can still replace more than 90\% of the manual work in generating the speaker-following subtitles for TV/movies.

\subsubsection{Better presentation of multimedia in noisy environments and for users with hearing impairment}
In noisy environments (e.g. subways and squares) or in a quiet public area, audio from the video may be hard to hear clearly or is muted. The viewer may then need speaker-following subtitles to know \enquote{who is speaking or what is being said} in order to better understand and enjoy the multimedia such as news, music videos, and television interviews.

As shown in the user study conducted by Hong et al. \citeyear{hong2010dynamic}, videos with dynamic subtitles can help users with hearing impairment understand the video contents. There are more than 66 million people suffering from hearing impairment, who can potentially benefit from the technology presented here.

\subsubsection{Automatic speech balloon layout for 2D/3D game avatars}
Previous methods \cite{park2008smart2,park2008smart} simply put speech balloons above game avatars' heads. Our system can be used for automatic word balloon positioning for avatars in 2D/3D games to provide better presentation of speech for game avatars.

\subsection{Limitations}
Our system cannot always correctly handle scenes where there are no speakers such as subtitles from a gramophone or a telephone message. If the speaker detection algorithm does not detect a speaker, then our system can still correctly put the subtitle into the default bottom screen position. However, if people in the scene move in such a way that confuses the speaker detection algorithm, our system may assign the subtitle to a wrong speaker, causing misunderstanding.

\section{Conclusion and Future Work}
\label{sec:conclusion}

To enhance video viewing experience, we have developed an automatic system to place video subtitles next to their corresponding speaker. The proposed system detects speakers and then computes optimal positions to place the subtitles. Improving the performance of both the speaker detection step and subtitle placement step enables a better presentation of video subtitles in TV and movies. A comprehensive usability study with 219 users validated the effectiveness of our system. We have also suggested several possible applications of our system.

Although our system consistently outperforms both static subtitling and a previous dynamic method, there is still room for improvement to increase user satisfaction in terms of overall viewing experience and eyestrain level. In this regard, we will further improve speaker detection accuracy especially in more challenging situations (e.g., the TV series \enquote{{\em ER}}, in which surgeons and nurses wear face masks and move around all the time). We also aim to extend our system to cope with sound-to-text libraries in online applications such as teleconferencing and real-time video streaming where no subtitle information is available.

\section{Acknowledgments}
The authors would like to thank all the participants of the usability study for their time and valuable feedback and comments.


\bibliographystyle{acmlarge}    
\bibliography{v2-acmlarge-sample}

\received{MM YYYY}{MM YYYY}{MM YYYY} 


\end{document}